\documentclass[aps,prb,twocolumn,groupedaddress,10pt]{revtex4-2}
\usepackage{amssymb}
\usepackage{graphicx}
\usepackage{color}
\usepackage{ulem}

\begin{document}

\title{High conductivity from cross-band electron pairing in flat-band systems}

\author{Maxim Trushin}
\affiliation{Department of Material Science and Engineering, National University of Singapore, 9 Engineering Drive 1, 117575, Singapore}
\affiliation{Institute for Functional Intelligent Materials, National University of Singapore, 4 Science Drive 2, 117544, Singapore}
\affiliation{Centre for Advanced 2D Materials, National University of Singapore, 6 Science Drive 2, 117546, Singapore}

\author{Liangtao Peng}
\affiliation{Department of Physics, National University of Singapore, 2 Science Drive 3, 117542, Singapore}
\affiliation{Centre for Advanced 2D Materials, National University of Singapore, 6 Science Drive 2, 117546, Singapore}

\author{Gargee Sharma}
\affiliation{School of Physical Sciences, Indian Institute of Technology Mandi, Mandi 175005, India}

\author{Giovanni Vignale}
\affiliation{Institute for Functional Intelligent Materials, National University of Singapore, 4 Science Drive 2, 117544, Singapore }
\affiliation{Centre for Advanced 2D Materials, National University of Singapore, 6 Science Drive 2, 117546, Singapore}
\affiliation{Department of Physics and Astronomy, University of Missouri, Columbia, Missouri 65211, USA}

\author{Shaffique Adam}
\affiliation{Department of Material Science and Engineering, National University of Singapore, 9 Engineering Drive 1, 117575, Singapore}
\affiliation{Yale-NUS College, 16 College Avenue West, 138527, Singapore}
\affiliation{Department of Physics, National University of Singapore, 2 Science Drive 3, 117542, Singapore}
\affiliation{Centre for Advanced 2D Materials, National University of Singapore, 6 Science Drive 2, 117546, Singapore}

\date{\today}

\begin{abstract}
Electrons in condensed matter may transition into a variety of broken-symmetry phase states due to electron-electron interactions.
Applying diverse mean-field approximations to the interaction term is arguably the simplest way to identify the phase states
theoretically possible in a given setting.
Here, we explore electron-electron attraction in a two-band system comprising symmetric conduction and valence bands touching each other
at a single point.
We assume a mean-field pairing between the electrons having opposite spins, momenta, and,
in contrast to the conventional superconducting pairing, residing in opposite bands, i.e., having opposite energies.
We show that electrons transition into a novel correlated ground state if and only if the bands are flat enough, i.e. 
the transition is impossible in the case of conventional parabolic bands.
Although this state is not superconducting in the usual sense
and does not exhibit a gap in its excitation spectrum, it is nevertheless immune to elastic scattering
caused by any kind of disorder, and is therefore expected to exhibit high electric conductivity at
low temperature, mimicking the behavior of a real superconductor.
Having in mind the recent experimental realizations of flat-band electronic systems in twisted multilayers,
we foresee an exciting opportunity for observing a new class of highly conductive materials.
\end{abstract}

\maketitle

\section{Introduction}

Free electrons are characterized by a finite rest mass, implying a quadratic relation between their energy and momentum.
When low-energy electrons trudge through a periodic structure, they still behave like free electrons
with an effective mass altered by the lattice potential.
In some exceptional limiting cases, 
the effective mass may vanish completely (graphene \cite{PR1947wallace,PRB2002ordejon}, topological insulator surfaces \cite{PRB2012mele-topological}, metallic nanotubes \cite{PRL1997nanotubes}) or acquire a nearly infinite value (heavy fermion \cite{heavy-fermion-review}
and kagome-type \cite{PRL2019kagome} materials, twisted graphene bilayers and multilayers \cite{bistritzer2011moire,cao_unconventional_2018}).
The latter examples are known as flat-band electronic systems
with prominent electron-electron correlation and topological effects due to the band touching, see Ref. [\onlinecite{flatbandreview2021}] for a recent review.
The absolute majority of recent publications devoted to the touching flat electronic bands \cite{peotta2015superfluidity,PRL2019rossi,PRL2020bernevig,PRL2021huber} is inspired by ongoing hunt for a high-temperature superconducting state \cite{balents2020superconductivity,torma2022superconductivity,tian2023evidence}.
However, the electron-electron pairing in flat bands does not necessarily lead to superconductivity \cite{boundstate2018,bandtouching2022}.
To give an example, rhombohedral-stacked multilayer graphene hosts a pair of flat bands
touching at zero energy, giving rise to nonsuperconducting electronic states such as correlated insulators \cite{han2024correlated}.
Here, we demonstrate a theoretical possibility for a nonsuperconductive (but nevertheless highly conductive) state of 
two-dimensional (2D) paired electrons driven by divergent density of states (DOS)
at the flat-band touching point.

\begin{figure}
 \includegraphics[width=\columnwidth]{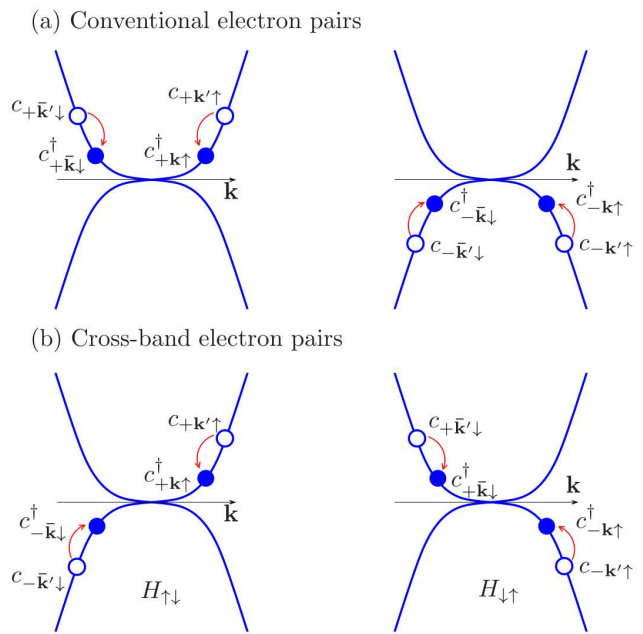}
 \caption{(a) Conventional and (b) cross-band electron pairing.
 The band dispersion is given by Eq. (\ref{dispersion}).
 Thin red arrows indicate the interaction processes involved in the reduced Hamiltonians.
 In the case of cross-band pairing, the terms $H_{\uparrow\downarrow}$ and $H_{\downarrow\uparrow}$ are given by
 Eqs. (\ref{Hupdown}) and (\ref{Hdownup}), respectively.
 Conventional pairing is time-reversal invariant, whereas our pairing is not.
 Nevertheless, our model as a whole is time-reversal invariant because $H_{\uparrow\downarrow}$ and $H_{\downarrow\uparrow}$
 are time-reversal partners of each other.
 The model is two-dimensional (2D), but only a given direction is depicted.
 Subscripts $+$ and $-$ designate the bands, while the negative of wave vector ${\bf k}$ is denoted by $\bar{\bf k}$. 
 \label{fig1}}
\end{figure}

In the case of a conventional electron pairing \cite{Tinkham1975}, the 
paired electrons have opposite spins and momenta while sitting at the same energy level.
The electrons interact with each other by exchanging momentum, Fig. \ref{fig1}(a).
The pairing mechanism may be due to
electron-phonon interaction, \cite{frohlich1952interaction} electron-electron screening, \cite{Kohn1965supercond} plasmon exchange, \cite{eliashberg1960interactions,grabowski1984superconductivity} spin fluctuations, \cite{spin-fluc1986} and other
phenomena \cite{DopingMottReview}.
Once the temperature is set below a certain critical value, the conventional electron pairs condense into a strongly correlated superconducting Bardeen-Cooper-Schrieffer (BCS) state  \cite{Tinkham1975}.
In this paper, we explore a pairing option shown in Fig. \ref{fig1}(b), 
where the two electrons having opposite momenta and spins are sitting at opposite energy levels,
as counted from the band-touching point. 
The model allows for the flatness tunability by means of the parameter $p\geq 2$, in accordance with the dispersion relation given by \cite{myPRB2019}
\begin{equation}
 \label{dispersion}
 \epsilon_k= \epsilon_0 \left(\frac{k}{k_0}\right)^p,
\end{equation}
where $k=|\mathbf{k}|$ is the absolute value of the wave vector, and $\epsilon_0$ can be interpreted as
the bandwidth of the flat region in the momentum subspace of size $k_0$. In contrast with other flat-band systems, here, 
$k_0$ does not extend over the entire Brillouin zone.  The electronic DOS is constant at $p=2$, and
it diverges at $p\to\infty$ as $1/\epsilon_k$, when the bands are perfectly flat at $k\leq k_0$.
The band flatness plays a crucial role in our model because the non-superconducting correlated ground state we have found exists
only if $p>2$ in 2D.
This contrasts with the conventional pairing that leads to the superconducting correlated ground state even at $p=2$.

Our cross-band pairing should not be confused with the conventional
interband pairing often associated with iron-based superconductors \cite{PRB2008,PRL2012,PRB2024}.
The interband pairing still occurs between electrons
having the same energy (Fermi energy) even though they may belong to different Fermi pockets.
When we are talking about cross-band pairing, we assume that the pairing occurs {\it across} the band-touching point.
This is the main difference between our cross-band pairing and what is usually considered interband pairing.

In what follows, we develop a many-body model using the cross-band pairing assumption within a mean-field approximation.
The conventional pairing is assumed to be blocked for one reason or another \cite{PRL2019rossi,PRL2020bernevig}.
We show that the cross-band electron pairing results in a correlated state, which we dubbed {\it frozen electrons}.
The paramount property of frozen electrons is that they turn out to be immune to elastic scattering.
As a consequence, we expect a resistivity drop upon transition into the correlated state that occurs at low temperatures.
We discuss possible realizations of the correlated state in twisted multilayer graphene samples, which are known to form
extremely flat bands and exhibit the nonsuperconducting low-temperature resistivity drop expected in our model.

\section{The model Hamiltonian}

The model Hamiltonian describing the pairing processes depicted in Fig. \ref{fig1}(b) can be written as $H=H_{\uparrow\downarrow}+H_{\downarrow\uparrow}$, where
\begin{eqnarray}
 \nonumber 
 H_{\uparrow\downarrow} & = &\sum_\mathbf{k} \left( \epsilon_k c_{+\mathbf{k}\uparrow}^\dagger c_{+\mathbf{k}\uparrow} -
 \epsilon_k c_{-\mathbf{\bar k}\downarrow}^\dagger c_{-\mathbf{\bar k}\downarrow} \right)\\
 \label{Hupdown} 
 & + &  \sum_{\mathbf{k}\mathbf{k}'} V_{\mathbf{k}\mathbf{k}'} c_{+\mathbf{k}\uparrow}^\dagger  c_{-\mathbf{\bar k}\downarrow}^\dagger c_{-\mathbf{\bar k}'\downarrow} c_{+\mathbf{k}'\uparrow},\\
 \nonumber 
 H_{\downarrow\uparrow} & = &\sum_\mathbf{k} \left( \epsilon_k c_{+\mathbf{\bar k}\downarrow}^\dagger c_{+\mathbf{\bar k}\downarrow} -
 \epsilon_k c_{-\mathbf{k}\uparrow}^\dagger c_{-\mathbf{k}\uparrow} \right)\\
 & + &  \sum_{\mathbf{k}\mathbf{k}'} V_{\mathbf{\bar k}\mathbf{\bar k}'} c_{+\mathbf{\bar k}\downarrow}^\dagger  c_{-\mathbf{k}\uparrow}^\dagger c_{-\mathbf{k}'\uparrow} c_{+\mathbf{\bar k}'\downarrow}.
\label{Hdownup}
\end{eqnarray}
Here, $\uparrow$ ($\downarrow$) is the spin-up (down) index, $+$ ($-$) is the conduction (valence) band index,
$\mathbf{\bar k}=-\mathbf{k}$, and $V_{\mathbf{k}\mathbf{k}'}$ is the Fourier transform of the pairing potential.
The two terms of the model Hamiltonian, $H_{\uparrow\downarrow}$ and $H_{\downarrow\uparrow}$,
are the time-reversal of each other that ensures the time-reversal invariance of the total $H$.
Each pair involves electrons having opposite momenta, spins, and band indices but
the same group velocity.
This is in stark contrast with the conventional pairing, 
$c_{\pm\mathbf{k}\uparrow}^\dagger  c_{\pm\mathbf{\bar k}\downarrow}^\dagger c_{\pm\mathbf{\bar k}'\downarrow} c_{\pm\mathbf{k}'\uparrow}$,
where the paired electrons are always at the same energy level despite having opposite spins and momenta, Fig. \ref{fig1}(a).
The conventional pairing is not included, and the pairing potential is not specified in our model Hamiltonian.

We now make a mean-field approximation \cite{flensberg2004many}  given by
\begin{eqnarray}
 \nonumber c_{+\mathbf{k}\uparrow}^\dagger  c_{-\mathbf{\bar k}\downarrow}^\dagger
 = && \langle c_{+\mathbf{k}\uparrow}^\dagger  c_{-\mathbf{\bar k}\downarrow}^\dagger\rangle
 + \left(c_{+\mathbf{k}\uparrow}^\dagger  c_{-\mathbf{\bar k}\downarrow}^\dagger
 -\langle c_{+\mathbf{k}\uparrow}^\dagger  c_{-\mathbf{\bar k}\downarrow}^\dagger\rangle\right),\\
 \nonumber  c_{-\mathbf{\bar k}'\downarrow} c_{+\mathbf{k}'\uparrow}  = &&
 \langle c_{-\mathbf{\bar k}'\downarrow} c_{+\mathbf{k}'\uparrow} \rangle \\
  && + \left(c_{-\mathbf{\bar k}'\downarrow} c_{+\mathbf{k}'\uparrow}-\langle c_{-\mathbf{\bar k}'\downarrow} c_{+\mathbf{k}'\uparrow}\rangle\right),\\
  \nonumber c_{+\mathbf{\bar k}\downarrow}^\dagger  c_{-\mathbf{k}\uparrow}^\dagger = && 
  \langle c_{+\mathbf{\bar k}\downarrow}^\dagger  c_{-\mathbf{k}\uparrow}^\dagger\rangle +
\left(c_{+\mathbf{\bar k}\downarrow}^\dagger  c_{-\mathbf{k}\uparrow}^\dagger-
\langle c_{+\mathbf{\bar k}\downarrow}^\dagger  c_{-\mathbf{k}\uparrow}^\dagger\rangle\right),\\
\nonumber c_{-\mathbf{k}'\uparrow} c_{+\mathbf{\bar k}'\downarrow}  =&&
c_{-\mathbf{k}'\uparrow} c_{+\mathbf{\bar k}'\downarrow} \\
&& + \left( c_{-\mathbf{k}'\uparrow} c_{+\mathbf{\bar k}'\downarrow} -\langle c_{-\mathbf{k}'\uparrow} c_{+\mathbf{\bar k}'\downarrow}\rangle \right),
\end{eqnarray}
where $\langle...\rangle$ represents an average value, and the terms in the brackets are assumed to be small.
The result reads
\begin{eqnarray}
 \nonumber H^{MF}_{\uparrow\downarrow} & = & \sum_\mathbf{k}
 (c_{+\mathbf{k}\uparrow}^\dagger, c_{- \mathbf{\bar k}\downarrow})
 \left(
\begin{array}{cc}
\epsilon_k & -\Delta_\mathbf{k}\\
-\Delta_\mathbf{k}^\dagger & \epsilon_k
\end{array}
 \right)
 \left(\begin{array}{c} c_{+\mathbf{k}\uparrow} \\
  c_{- \mathbf{\bar k}\downarrow}^\dagger \end{array}  \right)\\
\label{MF1} &+ & \sum_\mathbf{k} \left(\Delta_\mathbf{k}^\dagger \langle c_{-\mathbf{\bar k}\downarrow} c_{+\mathbf{k}\uparrow} \rangle - \epsilon_k \right),\\
\nonumber H^{MF}_{\downarrow\uparrow} & = & \sum_\mathbf{k}
 (c_{+\mathbf{\bar k}\downarrow}^\dagger, c_{-\mathbf{k}\uparrow})
 \left(
\begin{array}{cc}
\epsilon_k & -\Delta_\mathbf{k}\\
-\Delta_\mathbf{k}^\dagger & \epsilon_k
\end{array}
 \right)
 \left(\begin{array}{c} c_{+\mathbf{\bar k}\downarrow} \\
  c_{-\mathbf{k}\uparrow}^\dagger \end{array}  \right)\\
&+ & \sum_\mathbf{k} \left(\Delta_\mathbf{k}\langle c_{+ \mathbf{\bar k}\downarrow}^\dagger c_{-\mathbf{k}\uparrow}^\dagger \rangle - \epsilon_k \right),
\label{MF2}
\end{eqnarray}
where 
\begin{eqnarray}
\nonumber  \Delta_\mathbf{k} & = &-\sum_{\mathbf{k}'} V_{\mathbf{k}\mathbf{k}'} \langle c_{-\bar \mathbf{k}'\downarrow} c_{+\mathbf{k}'\uparrow} \rangle\\
\nonumber & = & -\sum_{\mathbf{k}'} V_{\mathbf{\bar k}\mathbf{\bar k}'} \langle c_{-\mathbf{k}'\uparrow} c_{+\mathbf{\bar k}'\downarrow} \rangle,\\
\nonumber \Delta_\mathbf{k}^\dagger 
& = & -\sum_{\mathbf{k}'} V_{\mathbf{k}\mathbf{k}'} \langle c_{+\mathbf{k}'\uparrow}^\dagger  c_{-\mathbf{\bar k}'\downarrow}^\dagger \rangle\\
& = & -\sum_{\mathbf{k}'} V_{\mathbf{\bar k}\mathbf{\bar k}'}  \langle c_{+\mathbf{\bar k}'\downarrow}^\dagger  c_{-\mathbf{k}'\uparrow}^\dagger \rangle
\label{Delta_k}
\end{eqnarray}
is the order parameter. Note the equal diagonal elements in $H^{MF}_{\uparrow\downarrow}$ and $H^{MF}_{\downarrow\uparrow}$.
Such a symmetry results in a Bogoljubov transformation with equal coherences and leads, as we show below, to the suppression of elastic electron scattering.

It is convenient to assume isotropic pairing and write the order parameter as $\Delta_\mathbf{k}=\Delta_k e^{i\delta_k}$.
Diagonalizing and transforming Eqs. (\ref{MF1}) and (\ref{MF2}) into the canonical form, we obtain
\begin{eqnarray}
\label{canon1} H^{MF}_{\uparrow\downarrow} & = &\sum_\mathbf{k} \Delta_k\left( 1+ e^{-i\delta_k} \langle c_{- \mathbf{\bar k}\downarrow} c_{+\mathbf{k}\uparrow} \rangle \right) \\
\nonumber  & + &\sum_\mathbf{k} \left[ \left(\epsilon_k-\Delta_k \right) \gamma_{0\mathbf{k}}^\dagger \gamma_{0\mathbf{k}}  +
           \left(-\epsilon_k-\Delta_k\right)\gamma_{1\mathbf{k}}^\dagger \gamma_{1\mathbf{k}} \right],\\
\label{canon2} H^{MF}_{\downarrow\uparrow} & = &\sum_\mathbf{k} \Delta_k\left( 1+ e^{i\delta_k} \langle c_{+\mathbf{\bar k}\downarrow}^\dagger c_{-\mathbf{k}\uparrow}^\dagger \rangle \right) \\
\nonumber  & + &\sum_\mathbf{k} \left[ \left(\epsilon_k-\Delta_k \right) \beta_{0\mathbf{k}}^\dagger \beta_{0\mathbf{k}}  +
           \left(-\epsilon_k-\Delta_k\right)\beta_{1\mathbf{k}}^\dagger \beta_{1\mathbf{k}} \right],           
\end{eqnarray}
where
\begin{eqnarray}
\nonumber  \langle c_{- \mathbf{\bar k}\downarrow} c_{+\mathbf{k}\uparrow} \rangle & = &
 \frac{1}{2}e^{i\delta_k}\left(\langle \gamma_{0\mathbf{k}}^\dagger \gamma_{0\mathbf{k}} \rangle + 
\langle \gamma_{1\mathbf{k}}^\dagger \gamma_{1\mathbf{k}} \rangle  -1 \right),\\
\nonumber  \langle c_{+\mathbf{\bar k}\downarrow}^\dagger c_{-\mathbf{k}\uparrow}^\dagger \rangle & = &
 \frac{1}{2}e^{-i\delta_k}\left(\langle \beta_{0\mathbf{k}}^\dagger \beta_{0\mathbf{k}} \rangle + 
\langle \beta_{1\mathbf{k}}^\dagger \beta_{1\mathbf{k}} \rangle  -1 \right)
\label{ave}
\end{eqnarray}
are the ground-state averages.
The quasiparticle creation and annihilation operators $\gamma_{0,1\mathbf{k}}^\dagger$ ($\beta_{0,1\mathbf{k}}^\dagger$)
and $\gamma_{0,1\mathbf{k}}$ ($\beta_{0,1\mathbf{k}}$) are related to the original electron operators via the same Bogoljubov transformation given by
\begin{eqnarray}
\nonumber \left(\begin{array}{c} c_{+\mathbf{k}\uparrow} \\
  c_{-\mathbf{\bar k}\downarrow}^\dagger \end{array}  \right)
 =P \left(\begin{array}{c} \gamma_{0\mathbf{k}} \\
\gamma_{1\mathbf{k}}^\dagger \end{array}  \right),
& \quad &
\left(\begin{array}{c} c_{+\mathbf{\bar k}\downarrow} \\
  c_{-\mathbf{k}\uparrow}^\dagger \end{array}  \right)=P
 \left(\begin{array}{c} \beta_{0\mathbf{k}} \\
\beta_{1\mathbf{k}}^\dagger \end{array}  \right),
\end{eqnarray}
where
\begin{equation}
 \label{bog}
 P = \frac{1}{\sqrt{2}}
 \left(
\begin{array}{cc}
1 & 1 \\
e^{-i\delta_k} & -e^{-i\delta_k}
\end{array}
 \right). 
\end{equation}
The order parameter phase $\delta_k$ does not play a role if no interface between the normal and correlated states is considered. We nevertheless retain $\delta_k$ for the sake of generality and future studies.

\section{Results}

As we shall see below, the order parameter equation has a solution if and only if the paring potential is attractive.
Hence, we assume a pointlike attractive interaction described by $V_{\mathbf{k}\mathbf{k}'}=-V_0$.
The order parameter can also be assumed to be $k$ independent, $\Delta_\mathbf{k}=\Delta e^{i\delta}$, and the order parameter equation reads
\begin{eqnarray}
\label{Delta-gamma} \Delta &= & \frac{V_0}{2}\sum_\mathbf{k}\left(\langle \gamma_{0\mathbf{k}}^\dagger \gamma_{0\mathbf{k}} \rangle + 
\langle \gamma_{1\mathbf{k}}^\dagger \gamma_{1\mathbf{k}} \rangle  -1 \right),\\
 \label{Delta-beta} \Delta &= & \frac{V_0}{2}\sum_\mathbf{k}\left(\langle \beta_{0\mathbf{k}}^\dagger \beta_{0\mathbf{k}} \rangle + 
\langle \beta_{1\mathbf{k}}^\dagger \beta_{1\mathbf{k}} \rangle  -1\right).
\end{eqnarray}
Note that the phase $e^{i\delta}$ is canceled out. Equations (\ref{Delta-gamma}) and (\ref{Delta-beta}) are equivalent, and the order parameter equation can be 
written in the final form as
\begin{equation}
 \label{gap-final}
 \Delta = \frac{V_0}{2}\sum_\mathbf{k}\left[n_F(\epsilon_k - \Delta) - n_F(\epsilon_k  + \Delta) \right], 
\end{equation}
where $n_F(\epsilon_k)=1/(1+e^{-\epsilon_k/T})$ is the Fermi-Dirac distribution, with $T$ being electron temperature. The Fermi energy level intersects 
the band-touching point and does not enter explicitly.
Like the conventional BCS gap equation \cite{Tinkham1975}, Eq. (\ref{gap-final}) 
is invariant under the substitution $\Delta\to -\Delta$ but has no solution if interactions are repulsive.
This is where the similarities end, and we show below that the properties of our correlated state have little to do with the BCS model.

\begin{figure}
 \includegraphics[width=\columnwidth]{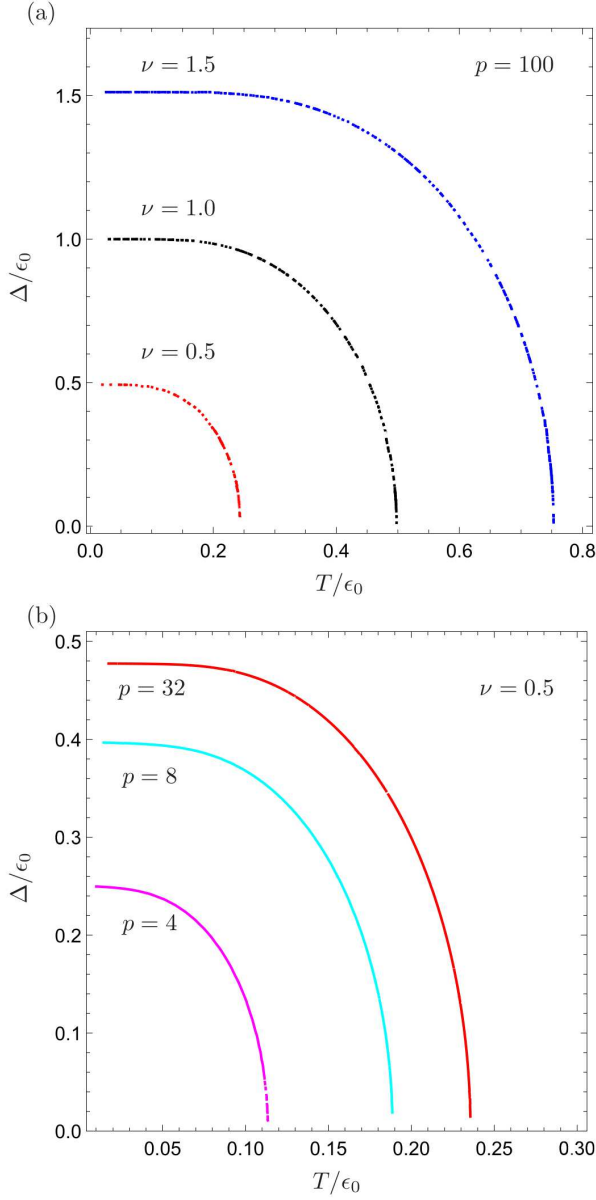}
 \caption{Solutions of the order parameter equation, Eq. (\ref{gap-2D}). 
(a) Solutions for different interaction strengths in the limit of extremely flat bands with $p=100$.
(b) Solutions for different band flatness parameter values at $\nu=0.5$.
Both the flatness and interaction strength make the critical temperature higher.
 \label{fig2}}
\end{figure}

Using the dispersion relation, Eq. (\ref{gap-final}) can be written in 2D explicitly as
\begin{equation}
 \label{gap-2D}
 \Delta = \frac{V_0k_0^2}{4\pi p}\int\limits_0^\infty\frac{d\epsilon}{\epsilon_0}\left(\frac{\epsilon}{\epsilon_0}\right)^{\frac{2-p}{p}}\left[n_F(\epsilon - \Delta) - n_F(\epsilon  + \Delta) \right].
\end{equation}
The integral can be written as the difference between two polylogarithms, but it is instructive to consider the limit $T=0$.
The order parameter then reads $\Delta=\epsilon_0\nu^{\frac{p}{p-2}}$, where
\begin{equation}
\label{nu}
\nu=\frac{V_0 k_0^2}{8\pi\epsilon_0}
\end{equation}
is the dimensionless interaction parameter. The order parameter increases with the band flatness,
and in the limit $p\to\infty$, we have $\Delta/\epsilon_0\to \nu$.
The finite-temperature solutions are shown in Fig. \ref{fig2}.
In addition to the function $\Delta(T)$, one can also figure out the critical temperature at which the order parameter vanishes for a given set of parameters.

The Bogoljubov transformation in Eq. (\ref{bog}) is like the one diagonalizing the BCS Hamiltonian \cite{flensberg2004many},
but the coherence factors, $u_k$ and $v_k$, have equal magnitude. The many-body ground state of Hamiltonian $H^{MF}=H^{MF}_{\uparrow\downarrow}+H^{MF}_{\downarrow\uparrow}$
must therefore have a similar structure. Introducing $k_\Delta$ (which satisfies $\epsilon_{k_\Delta}=\Delta$),
we find that the ground state can be written as
$|G\rangle = |G\rangle_{\uparrow\downarrow}\otimes|G\rangle_{\downarrow\uparrow}$, where
\begin{eqnarray} 
\nonumber && |G\rangle_{\uparrow\downarrow} = \prod\limits_{k\leq k_\Delta}\frac{1}{\sqrt{2}}\left(1 + e^{i\delta} c_{+\mathbf{k}\uparrow}^\dagger c_{-\mathbf{\bar k}\downarrow}^\dagger\right) \prod\limits_{k>k_\Delta} c_{-\mathbf{\bar k}\downarrow}^\dagger |0\rangle,\\
\nonumber && |G\rangle_{\downarrow\uparrow} = \prod\limits_{k\leq k_\Delta}
\frac{1}{\sqrt{2}}\left(1 + e^{i\delta} c_{+\mathbf{\bar k}\downarrow}^\dagger c_{-\mathbf{k}\uparrow}^\dagger\right)
\prod\limits_{k>k_\Delta} c_{-\mathbf{k}\uparrow}^\dagger |0 \rangle. \\
\label{M}
\end{eqnarray}
Indeed, taking the limit $T\to 0$ in Eqs. (\ref{MF1}) and (\ref{MF2}), the eigenvalues read
\begin{eqnarray}
 \nonumber \langle G|H^{MF}_{\uparrow\downarrow}|G\rangle_{\uparrow\downarrow}&& = \langle G|H^{MF}_{\downarrow\uparrow}|G\rangle_{\downarrow\uparrow}\\
 && =-\sum\limits_{k\leq k_\Delta}\frac{\Delta_k}{2}-\sum\limits_{k>k_\Delta}\epsilon_k.
\end{eqnarray}
The same can be deduced from the canonical Eqs. (\ref{canon1}) and (\ref{canon2}), and the total 
ground-state energy can be written as
\begin{equation}
 E_G=-\sum\limits_{k\leq k_\Delta}\Delta_k - 2\sum\limits_{k>k_\Delta}\epsilon_k.
\end{equation}

The ground-state function $|G\rangle$ must be contrasted with the one of a normal state. It is simply given by
\begin{equation}
 |N\rangle =\prod\limits_\mathbf{k} c_{-\mathbf{\bar k}\downarrow}^\dagger c_{-\mathbf{k}\uparrow}^\dagger |0 \rangle,
\end{equation}
and the energy reads
\begin{equation}
 E_N=- 2\sum\limits_k \epsilon_k.
\end{equation}
Obviously, $E_G$ and $E_N$ both diverge to negative infinity. However, the difference $E_G-E_N$ is well defined and can
be written in 2D explicitly as
\begin{eqnarray}
E_G-E_N & = & -\sum\limits_{k\leq k_\Delta}\left(\Delta_k-2\epsilon_k\right)\\
&= & -\frac{k_0^2}{2\pi p}\int\limits_0^\Delta\frac{d\epsilon}{\epsilon_0}\left(\frac{\epsilon}{\epsilon_0}\right)^{\frac{2-p}{p}}\left(\Delta -2\epsilon\right)\\
& = & -\frac{\Delta k_0^2}{2\pi}\left(\frac{\Delta}{\epsilon_0}\right)^{\frac{2}{p}}\left(\frac{1}{2}-\frac{2}{2+p}\right)\\
& = & - \frac{\epsilon_0 k_0^2}{4\pi}\frac{p-2}{p+2}\nu^\frac{p+2}{p-2}.
\label{DeltaE}
\end{eqnarray}
Thus, $E_G < E_N$ if $p>2$, i.e., $|G\rangle$ is indeed the ground state of $H$.
The ground-state energy level dives deeper below the normal-state energy with increasing band flatness.
However, $E_G \geq  E_N $ at $p\leq 2$, i.e., the ground state is normal (not correlated) if the band flatness is not strong enough.
This is the main result of this paper indicating a possibility of a novel ground state for electrons near the flat-band touching point.

\begin{figure*}
 \includegraphics[width=\textwidth]{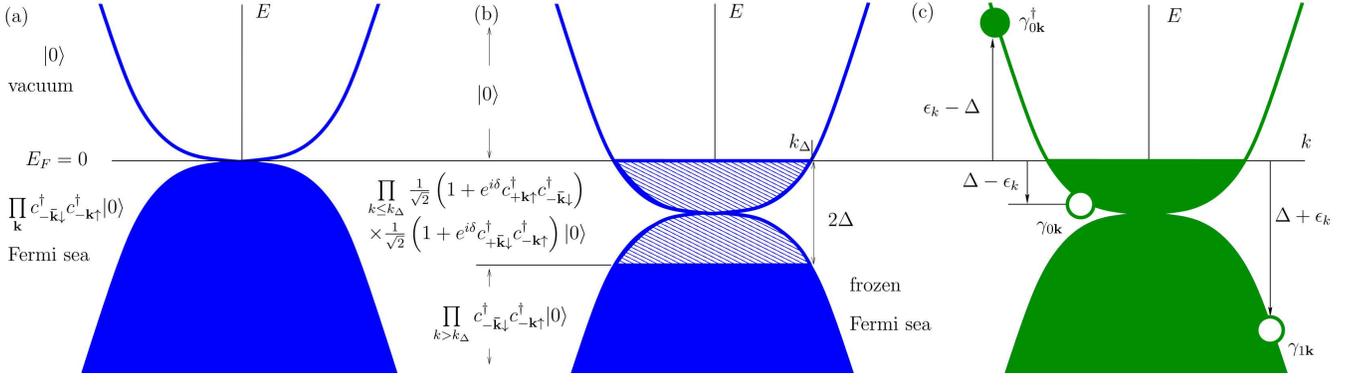}
 \caption{ (a) The normal electron state with the Fermi energy intercepting the flat-band touching point ($E_F=0$).
 (b) The correlated state splits into two sectors with $k \leq k_\Delta$ and $k > k_\Delta$, see Eq. (\ref{M}).
 The former consists of strongly correlated electrons, whereas the latter represents normal electrons, and the two sectors together 
 can be seen as a layer of frozen electrons floating on a Fermi liquid.
 The thickness of the frozen layer is $2\Delta$, with $\Delta$ being the self-consistent order parameter.
 (c) The correlated state and excitations in terms of the quasiparticle operators $\gamma_{0{\mathbf k}}$ and $\gamma_{1{\mathbf k}}$.
 The $\gamma_{1{\mathbf k}}$-excitations are gapped by $\Delta$ even though the overall spectrum is gapless.
 The same is true for $\beta_{0{\mathbf k}}$ and $\beta_{1{\mathbf k}}$.
 \label{fig3}}
\end{figure*}

The model can also be adapted to 1D or 3D electrons.
One can easily derive an equation like Eq. (\ref{DeltaE}) and see that
the correlated ground state is possible at $p>3$ in the 3D case, and at $p>1$ is a 1D limit.
Since the absolute majority of the flat-band electronic systems are 2D, we focus on a 2D case here and in what follows.

\section{Discussion}

The difference between the normal and correlated electron states is demonstrated in Figs. \ref{fig3}(a) and \ref{fig3}(b).
Upon transition from the normal to the correlated state, the electrons near the flat-band touching point are instantly redistributed 
within the energy interval $[-\Delta,+\Delta]$ so that the electronic states acquire fractional occupation in momentum space.
The states outside the interval remain normal, i.e., fully occupied below the level $-\Delta$ 
and completely empty above $+\Delta$. We think about the correlated electrons as {\it frozen} ones because  
of the similarity between a frozen surface layer of real water and the layer of paired electrons near the Fermi surface in momentum space.
The similarity can go even further with respect to perturbations. Obviously, it is hard to disturb the surface of frozen water,
as one must penetrate through the ice layer. We predict that it is also hard to disturb the electrons covered by the frozen layer.

To test this prediction, we consider electron transport. The disorder scattering Hamiltonian can be written as
\begin{eqnarray}
\nonumber H_\mathrm{dis} & = &\sum\limits_{\mathbf{kk}'} U_{{\mathbf k}{\mathbf k}'} 
\left( c_{+{\mathbf k}'\uparrow}^\dagger c_{+{\mathbf k}\uparrow} + c_{-\mathbf{\bar k}'\downarrow}^\dagger c_{-\mathbf{\bar k}\downarrow} 
+ c_{+\mathbf{\bar k}'\downarrow}^\dagger c_{+\mathbf{\bar k}\downarrow} \right.\\
\nonumber && \left.  +c_{-{\mathbf k}'\uparrow}^\dagger c_{-{\mathbf k}\uparrow}
+c_{+{\mathbf k}'\uparrow}^\dagger c_{-{\mathbf k}\uparrow} + c_{-{\mathbf k}'\uparrow}^\dagger c_{+{\mathbf k}\uparrow} \right.\\
&& \left. + c_{-\mathbf{\bar k}'\downarrow}^\dagger c_{+\mathbf{\bar k}\downarrow} + c_{+\mathbf{\bar k}'\downarrow}^\dagger c_{-\mathbf{\bar k}\downarrow}
\right).
\label{Hdis}
\end{eqnarray}
Here, $U_{{\mathbf k}{\mathbf k}'}=U_{{\mathbf k}'{\mathbf k}}$ is the matrix element of a smooth potential unable to induce spin-flip and intervalley scattering.
Equation (\ref{Hdis}) can be rewritten in terms of the quasiparticle operators according to the Bogoljubov transformation in Eq. (\ref{bog}) as
\begin{eqnarray}
\nonumber H_\mathrm{dis}  & = & \sum\limits_{\mathbf{kk}'} U_{{\mathbf k}{\mathbf k}'} \left(
 \gamma_{0\mathbf{k}'}^\dagger \gamma_{1\mathbf{k}}^\dagger + \gamma_{1\mathbf{k}'} \gamma_{0\mathbf{k}}
 +\beta_{0\mathbf{k}'}^\dagger \beta_{1\mathbf{k}}^\dagger + \beta_{1\mathbf{k}'} \beta_{0\mathbf{k}}
  \right.\\
 \nonumber && \left. +e^{i\delta}\beta_{1\mathbf{k}'}\gamma_{0\mathbf{k}}^\dagger+ e^{i\delta}\gamma_{1\mathbf{k}'} \beta_{0\mathbf{k}}^\dagger  
 + e^{-i\delta}\beta_{0\mathbf{k}'}\gamma_{1\mathbf{k}}^\dagger \right. \\
 && \left. + e^{-i\delta}\gamma_{0\mathbf{k}'}\beta_{1\mathbf{k}}^\dagger
 \right),
 \label{Hdis2}
\end{eqnarray}
see Appendix \ref{appA}.

Figure \ref{fig3}(c) demonstrates that the low-energy excitations are created by quasiparticle operators of ``0'' type.
In contrast, the excitations of type ``1'' are gapped.
However, all terms in Eq. (\ref{Hdis2}) contain one quasiparticle operator of type ``1''.
Hence, calculating the low-energy scattering matrix elements, we obtain 
an expectation value of a quartic product in which three quasiparticle operators 
are always of ``0''-type
and one operator of type ``1''. The expectation value must necessarily vanish since the numbers
of ``0'' and ``1''-type operators are not balanced.
The scattering becomes only possible when it is accompanied by substantial electron energy change larger than $\Delta$.
Alternatively, we would allow for the interband electronic transitions with spin flip.
However, spin flip requires magnetic impurities which are usually in much lower concentrations 
than nonmagnetic ones.

The frozen-electron effect could possibly be detected in low-temperature electrical resistivity measurements.
If the temperature drops below the critical one (see Fig. \ref{fig2}), then the electrons transition into the correlated ground state.
While the electrons are in a frozen state, they are immune to elastic scattering; hence, they do not feel charged impurities and other defects.
One is only left with magnetic impurities, if any, and inelastic scatterers, like phonons, which are much weaker at low temperatures. 
As a consequence, we expect a sudden drop in electrical resistivity below a certain critical temperature.

To substantiate this claim, we calculate electrical conductivity in our correlated state.
To do that, we follow Ref. \cite{Parks1969} and solve the Boltzmann equation for each quasiparticle type
$\kappa=\{\gamma_{0,1}, \beta_{0,1}\}$ given by
\begin{equation}
 \label{Boltz}
 \frac{\partial f_{\kappa \mathbf{k}}}{\partial t} + \mathbf{v}\frac{\partial f_{\kappa \mathbf{k}}}{\partial \mathbf{r}}
 + \frac{e}{\hbar}{\cal\bf E} \frac{\partial f_{\kappa \mathbf{k}}}{\partial \mathbf{k}} = 
 \left(\frac{\partial f_{\kappa \mathbf{k}}}{\partial t}\right)_\mathrm{coll}.
 \end{equation}
We assume a steady state ($f_{\kappa \mathbf{k}}$ does not depend on $t$) without spatial inhomogeneities 
($f_{\kappa \mathbf{k}}$ does not depend on $\mathbf{r}$), a linear response regime (small homogeneous  electric field 
$|{\cal\bf E}|={\cal E}_x$ along the $x$ axis),
and short-range elastic scattering on impurities described within the relaxation-time approximation.
The solution of Eq. (\ref{Boltz}) can then be written as 
$f_{\kappa \mathbf{k}} = f_{\kappa \mathbf{k}}^{(0)} + f_{\kappa \mathbf{k}}^{(1)}$, where
$f_{\kappa \mathbf{k}}^{(0)}$ is $f_{\kappa \mathbf{k}}$ at ${\cal E}_x=0$, and
\begin{equation}
 \label{Boltz2}
 f_{\kappa \mathbf{k}}^{(1)} = e{\cal E}_x v_x \tau_{\kappa k}
 \left(-\frac{\partial f_{\kappa \mathbf{k}}^{(0)}}{\partial \epsilon_k} \right),
\end{equation}
where $\hbar v_x=\partial \epsilon_k/\partial k_x$, and electrical conductivity reads
\begin{equation}
 \sigma_{xx}= \frac{e}{\cal E}_x\sum\limits_\kappa\int\frac{d^2 k}{(2\pi)^2}
 v_xf_{\kappa \mathbf{k}}^{(1)}.
\end{equation}
The conductivity is mostly determined by the momentum relaxation time $\tau_{\kappa k}$ considered in Appendix \ref{appA}.
We consider all the scattering channels specified by the disorder Hamiltonian in Eq. (\ref{Hdis2})
and find that they do not contribute to $\tau_{\kappa k}$ due to the energy conservation.
The situation is like the interband scattering channels forbidden in the normal state
if electron-disorder scattering is elastic. 
In the normal state, however, there are always intraband scattering channels which make
$\tau_{\kappa k}$ finite, see Appendix \ref{appA}.

In fact, the low-energy quasiparticle states are protected from scattering by the symmetry of the Bogoljubov transformation having equal coherences. The protection effect is geometric or topological in some sense. 
Hence, the band symmetry with respect to the touching point plays a crucial role.
The symmetry is broken once the conduction and valence band dispersions are not the same or the Fermi level does not intercept the band touching point.
In either case, the Bogoljubov transformation is not symmetric, and the electron scattering terms do not cancel out in $H_\mathrm{dis}$.

Indeed, if the conduction band dispersion $\epsilon_k^v$ does not equal the valence one $\epsilon_k^c$, then $H^{MF}_{\uparrow\downarrow}$ takes the form:
\begin{eqnarray}
 \nonumber H^{MF}_{\uparrow\downarrow} & = & \sum_\mathbf{k}
 (c_{+\mathbf{k}\uparrow}^\dagger, c_{- \mathbf{\bar k}\downarrow})
 \left(
\begin{array}{cc}
\xi_k & -\Delta_\mathbf{k}\\
-\Delta_\mathbf{k}^\dagger & -\xi_k
\end{array}
 \right)
 \left(\begin{array}{c} c_{+\mathbf{k}\uparrow} \\
  c_{- \mathbf{\bar k}\downarrow}^\dagger \end{array}  \right)\\
 \nonumber  &+ & \sum_\mathbf{k} \frac{\epsilon_k^c + \epsilon_k^v}{2}\left(c_{+\mathbf{k}\uparrow}^\dagger c_{+\mathbf{k}\uparrow}
- c_{- \mathbf{\bar k}\downarrow}^\dagger c_{- \mathbf{\bar k}\downarrow}\right)\\
&+ & \sum_\mathbf{k} \left(\Delta_\mathbf{k}^\dagger \langle c_{-\mathbf{\bar k}\downarrow} c_{+\mathbf{k}\uparrow} \rangle + 
\xi_k\right),
\label{MFasym} 
\end{eqnarray}
where $\xi_k=(\epsilon_k^c - \epsilon_k^v)/2$.
The canonical form of the Hamiltonian then reads
\begin{eqnarray}
\nonumber  H^{MF}_{\uparrow\downarrow} & = &\sum_\mathbf{k}\left(\Delta_k e^{-i\delta_k} \langle c_{- \mathbf{\bar k}\downarrow} c_{+\mathbf{k}\uparrow} \rangle  + \xi_k +\sqrt{\xi_k^2 + \Delta_k^2} \right) \\
\nonumber  & + &\sum_\mathbf{k} \left[ \left(\frac{\epsilon_k^c+\epsilon_k^v}{2}-\sqrt{\Delta_k^2+\xi_k^2} \right) \gamma_{0\mathbf{k}}^\dagger \gamma_{0\mathbf{k}}  \right. \\
&& +    \left. \left(-\frac{\epsilon_k^c+\epsilon_k^v}{2}-\sqrt{\Delta_k^2+\xi_k^2}\right)\gamma_{1\mathbf{k}}^\dagger \gamma_{1\mathbf{k}} \right],
\end{eqnarray}
where the order parameter $\Delta_\mathbf{k}=\Delta e^{i\delta}$ can be computed from the equation given by
\begin{eqnarray}
 \nonumber \Delta &= & \frac{V_0}{2}\sum_\mathbf{k}\frac{\Delta}{\sqrt{\Delta^2+\xi_k^2}}\left(\langle \gamma_{0\mathbf{k}}^\dagger \gamma_{0\mathbf{k}} \rangle + 
\langle \gamma_{1\mathbf{k}}^\dagger \gamma_{1\mathbf{k}} \rangle  -1 \right).\\
\label{Delta-gamma2}
\end{eqnarray}
In contrast with Eq. (\ref{bog}), the Bogoljubov transformation is given by the matrix: 
\begin{equation}
  \label{bog2}
 P =  \left(
\begin{array}{cc}
 \sin\frac{\theta_k}{2} & -\cos\frac{\theta_k}{2} \\
\cos\frac{\theta_k}{2}e^{-i\delta} & \sin\frac{\theta_k}{2}e^{-i\delta}
\end{array}
 \right),
\end{equation}
where $\tan\theta_k=\Delta/\xi_k$. Obviously, the scattering terms of the form 
$\gamma_{0\mathbf{k}'}^\dagger \gamma_{0\mathbf{k}}$ do not cancel out in the disorder Hamiltonian $H_\mathrm{dis}$,
and electronic resistivity is not reduced.
However, an asymmetry in the Bogoljubov transformation in Eq. (\ref{bog2}) can be somewhat suppressed 
by stronger interactions so that $\Delta\gg \xi_k$ and $\theta_k \to \pi/2$. 

The symmetry of the Bogoljubov transformation in Eq. (\ref{bog}) can also be broken by a finite Fermi energy level ($E_F\neq 0$).
Note that the chemical potential is equivalent to the Fermi energy
in the zero-temperature limit. Hence, considering finite Fermi energy makes it possible
to draw conclusions regarding the finite doping effects on our frozen electron state.
The mean-field Hamiltonian then takes the form given by
\begin{eqnarray}
 \nonumber H^{MF}_{\uparrow\downarrow} & = & \sum_\mathbf{k}
 (c_{+\mathbf{k}\uparrow}^\dagger, c_{-\bar \mathbf{k}\downarrow})
 \left(
\begin{array}{cc}
\epsilon_k-E_F & -\Delta_\mathbf{k}\\
-\Delta_\mathbf{k}^\dagger & \epsilon_k + E_F
\end{array}
 \right)
 \left(\begin{array}{c} c_{+\mathbf{k}\uparrow} \\
  c_{-\bar \mathbf{k}\downarrow}^\dagger \end{array}  \right)\\
&+ & \sum_\mathbf{k} \left(\Delta_\mathbf{k}^\dagger \langle c_{-\bar \mathbf{k}\downarrow} c_{+\mathbf{k}\uparrow} \rangle 
- \epsilon_k - E_F \right).
\label{MF-EF}
\end{eqnarray}
The canonical form of the Hamiltonian reads
\begin{eqnarray}
H^{MF}_{\uparrow\downarrow} & = &\sum_\mathbf{k} \left( \Delta_k e^{-i\delta_k} \langle c_{-\bar \mathbf{k}\downarrow} c_{+\mathbf{k}\uparrow} \rangle + \Delta_F - E_F \right) \\
\nonumber  & + &\sum_\mathbf{k} \left[ \left(\epsilon_k-\Delta_F \right) \gamma_{0\mathbf{k}}^\dagger \gamma_{0\mathbf{k}}  +
           \left(-\epsilon_k-\Delta_F\right)\gamma_{1\mathbf{k}}^\dagger \gamma_{1\mathbf{k}} \right],
\end{eqnarray}
where  $\Delta_F=\sqrt{\Delta_k^2+E_F^2}$ is a new $E_F$-dependent order parameter.
Assuming that $\Delta_\mathbf{k}=\Delta e^{i\delta}$ is independent of $k$, the order parameter equation can be written as
\begin{eqnarray}
\label{Delta-gamma3} 
\Delta_F &= & \frac{V_0}{2}\sum_\mathbf{k}\left(\langle \gamma_{0\mathbf{k}}^\dagger \gamma_{0\mathbf{k}} \rangle + 
\langle \gamma_{1\mathbf{k}}^\dagger \gamma_{1\mathbf{k}} \rangle  -1 \right).
\end{eqnarray}
The final order parameter equation can also be obtained from Eq. (\ref{gap-final}) by the substitution $\Delta \to \Delta_F$.
However, the Bogoljubov transformation is asymmetric and given by 
\begin{equation}
  \label{bog3}
 P =  \left(
\begin{array}{cc}
  \cos\frac{\zeta}{2} & \sin\frac{\zeta}{2} \\
\sin\frac{\zeta}{2}e^{-i\delta} & -\cos\frac{\zeta}{2}e^{-i\delta}
\end{array}
 \right),
\end{equation}
where $\tan\zeta=\Delta/E_F$. 
The asymmetry does not allow for the cancellation of scattering terms in $H_\mathrm{dis}$ 
but again can be suppressed by stronger interactions making $\Delta \gg E_F$.

\begin{figure}
 \includegraphics[width=\columnwidth]{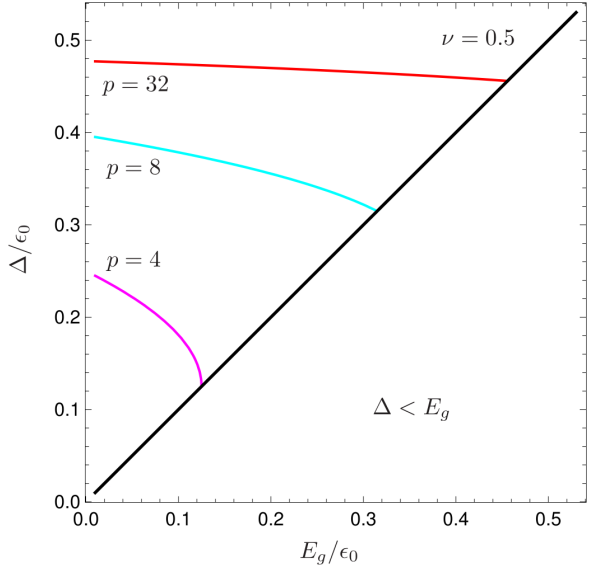}
 \caption{Solutions of the order parameter Eq. (\ref{gap-4-T0}) for different band flatness parameter values 
 in the presence of a band gap $E_g$. The color code is the same as in Fig. \ref{fig2}(b).
 Note that the order parameter depends weakly on the band gap if $p\gg 1$.
 The region below the diagonal is excluded, as the correlated state is not a ground state at $\Delta<E_g$
 ($E_G-E_N>0$), see Eq. (\ref{final}).
 \label{fig4}}
\end{figure}

A finite band gap does not break the symmetry of the Bogoljubov transformation in Eq. (\ref{bog}),
but it changes the order parameter equation as well as the ground-state energy.
Indeed, introducing a gapped dispersion $\epsilon_k^g=\epsilon_k+E_g/2$,
we can easily write the canonical mean-field Hamiltonian as
\begin{eqnarray}
\label{canon4} H^{MF}_{\uparrow\downarrow} & = &\sum_\mathbf{k} \Delta_k\left( 1+ e^{-i\delta_k} \langle c_{- \mathbf{\bar k}\downarrow} c_{+\mathbf{k}\uparrow} \rangle \right) \\
\nonumber  & + &\sum_\mathbf{k} \left[ \left(\epsilon_k^g-\Delta_k \right) \gamma_{0\mathbf{k}}^\dagger \gamma_{0\mathbf{k}}  +
           \left(-\epsilon_k^g-\Delta_k\right)\gamma_{1\mathbf{k}}^\dagger \gamma_{1\mathbf{k}} \right].
\end{eqnarray}
The order parameter equation at $\Delta_\mathbf{k}=\Delta e^{i\delta}$ then reads
\begin{eqnarray}
 \nonumber  \Delta & =& \frac{V_0k_0^2}{4\pi p}\int\limits_0^\infty\frac{d\epsilon}{\epsilon_0}\left(\frac{\epsilon}{\epsilon_0}\right)^{\frac{2-p}{p}}\left[n_F(\epsilon+E_g/2 - \Delta) \right.\\
 && \left. - n_F(\epsilon+E_g/2  + \Delta) \right],
 \label{gap-4}
\end{eqnarray}
which at $T=0$ takes the form:
\begin{equation}
 \label{gap-4-T0}
 \frac{\Delta}{\epsilon_0}=\nu \left(\frac{\Delta}{\epsilon_0}-\frac{E_g}{2\epsilon_0}\right)^{\frac{2}{p}},
\end{equation}
where $\nu$ is given by Eq. (\ref{nu}).
The energy difference between the correlated and normal states then reads
\begin{eqnarray}
\nonumber E_G-E_N &= & -\frac{k_0^2}{2\pi p}\int\limits_0^{\Delta-E_g/2}\frac{d\epsilon}{\epsilon_0}\left(\frac{\epsilon}{\epsilon_0}\right)^{\frac{2-p}{p}}\left(\Delta-E_g -2\epsilon\right)\\
& = & -\frac{k_0^2}{4\pi}\frac{\Delta(\Delta-E_g)}{\nu \epsilon_0}\frac{p-2}{p+2}.
\label{final}
\end{eqnarray}
Thus, the correlated state is a ground state if and only if $p>2$ {\it and} interactions are so strong that $\Delta>E_g$.
Solutions of the order parameter equation are shown in Fig. \ref{fig4} at $E_g>0$ and $T=0$.
The region $\Delta<E_g$ is excluded even though the formal solutions still exist.
The finite band gap reduces $\Delta$ and may even switch the state back to the normal one,
especially if the flatness is not strong enough.
Hence, the effects of a finite band gap are not desirable but can be suppressed by strong interactions and band flatness.

\begin{figure}
 \includegraphics[width=\columnwidth]{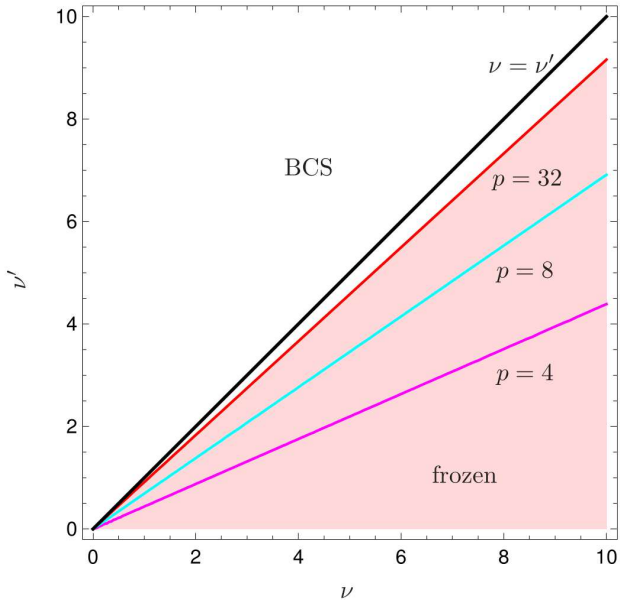}
 \caption{Phase diagram demonstrating the conventional BCS and our frozen electron states depending on
 the interaction parameters, $\nu$ (cross-band) and $\nu'$ (BCS-like), given by Eqs. (\ref{nu}) and (\ref{nu1}), respectively.
 The region with $\nu'>\nu$ always represents a BCS state. At $\nu'<\nu$, however, 
 the border separating the two states is determined by the band flatness parameter $p$.
  The color code of the lines is the same as in Fig. \ref{fig2}(b). The shaded area corresponds to the region of frozen electrons with $p=32$. 
 \label{fig5}}
\end{figure}

Finally, we consider the conventional pairing mentioned below Eq. (\ref{Hdownup}) 
and figure out the critical interaction strength at which the BCS ground-state energy is lower than $E_G$ given by Eq. (\ref{DeltaE}).
The BCS mean-field Hamiltonian can be written as
\begin{eqnarray}
 \nonumber H^{MF}_\mathrm{BCS} & = &\sum_\mathbf{k} \left[\Delta'_\mathbf{k}\left(\langle c_{+ \mathbf{\bar k}\downarrow} c_{+\mathbf{k}\uparrow} \rangle 
  + \langle c_{- \mathbf{\bar k}\downarrow} c_{-\mathbf{k}\uparrow} \rangle\right)- 2E_k \right]\\
\nonumber  & + &\sum_\mathbf{k} E_k \left(\gamma_{0\mathbf{k}}^\dagger \gamma_{0\mathbf{k}}  +
           \gamma_{1\mathbf{k}}^\dagger \gamma_{1\mathbf{k}} +\beta_{0\mathbf{k}}^\dagger \beta_{0\mathbf{k}}
           + \beta_{1\mathbf{k}}^\dagger \beta_{1\mathbf{k}}\right),\\
 \label{BCSMF}
\end{eqnarray}
where the quasiparticle operators  $\gamma_{0,1\mathbf{k}}$ and $\beta_{0,1\mathbf{k}}$ are related 
to electronic ones through the well-known Bogoljubov transformation \cite{flensberg2004many}, 
$E_k=\sqrt{\Delta'^2_k+\epsilon_k^2}$  is the quasiparticle excitation spectrum, and
the order parameter reads
\begin{eqnarray}
  \Delta'_\mathbf{k} & = &-\sum_{\mathbf{k}'} V'_{\mathbf{k}\mathbf{k}'} \langle c_{\pm\bar \mathbf{k}'\downarrow} c_{\pm\mathbf{k}'\uparrow} \rangle,\\
 \Delta'^\dagger_\mathbf{k} 
& = & -\sum_{\mathbf{k}'} V'_{\mathbf{k}\mathbf{k}'} \langle c_{\pm\mathbf{k}'\uparrow}^\dagger  c_{\pm\mathbf{\bar k}'\downarrow}^\dagger \rangle
\label{Delta_BCS}
\end{eqnarray}
with $V'_{\mathbf{k}\mathbf{k}'}$ being the BCS (intraband) pairing potential.
To proceed, we set $V'_{\mathbf{k}\mathbf{k}'}=V'_0$ and define the dimensionless interaction parameter as 
\begin{equation}
\label{nu1}
\nu'=\frac{V'_0 k_0^2}{8\pi\epsilon_0},
\end{equation}
cf. Eq. (\ref{nu}). In the limit $T=0$ the order parameter explicitly reads
\begin{equation}
 \Delta'=\epsilon_0\left[\frac{\nu'}{p\sqrt{\pi}} \Gamma\left(\frac{1}{2}-\frac{1}{p}\right) 
 \Gamma\left(\frac{1}{p}\right)\right]^{\frac{p}{p-2}},
\end{equation}
where $\Gamma(x)$ is the Gamma function. 
The ground-state energy counted from the normal-state energy level can be written as
\begin{eqnarray}
E'_G-E_N & = & -\sum\limits_{k}\left(2\epsilon_k-2E_k +\frac{\Delta'^2}{E_k}\right)\\
\nonumber &= & \frac{k_0^2}{2\pi p}\int\limits_0^\infty\frac{d\epsilon}{\epsilon_0}\left(\frac{\epsilon}{\epsilon_0}\right)^{\frac{2-p}{p}}\left(2\epsilon-2\sqrt{\epsilon^2 + \Delta'^2} \right. \\
&& \left. +\frac{\Delta'^2}{\sqrt{\epsilon^2 + \Delta'^2}}\right)\\
\nonumber & = & \frac{\epsilon_0 k_0^2}{4p\pi^{\frac{3}{2}}}\left(\frac{\Delta'}{\epsilon_0}\right)^{\frac{2+p}{p}}\Gamma\left(\frac{1}{p}\right) \\
& & \times \left[\Gamma\left(-\frac{1}{2}-\frac{1}{p}\right)+\Gamma\left(\frac{1}{2}-\frac{1}{p}\right) \right],
\label{DeltaE1}
\end{eqnarray}
cf. Eq. (\ref{DeltaE}).
The difference $E'_G-E_N$ is always negative, but $E'_G-E_G$ may be positive or negative
depending on the interaction parameters $\nu$, $\nu'$ and band flatness $p$.
The condition $E'_G-E_G=0$ can be explicitly written as
\begin{eqnarray}
\nonumber && \frac{1}{p\sqrt{\pi}}\left[\frac{\nu'}{p\sqrt{\pi}} \Gamma\left(\frac{1}{2}-\frac{1}{p}\right) 
 \Gamma\left(\frac{1}{p}\right)\right]^{\frac{p+2}{p-2}}\Gamma\left(\frac{1}{p}\right) \\
\nonumber & & \times \left[\Gamma\left(-\frac{1}{2}-\frac{1}{p}\right)+\Gamma\left(\frac{1}{2}-\frac{1}{p}\right) \right]
+\frac{p-2}{p+2} \nu^{\frac{p+2}{p-2}}=0.\\
\label{DeltaEG1}
\end{eqnarray}
The solution of Eq. (\ref{DeltaEG1}) is shown in Fig. \ref{fig5} for different values of $p$.
The regions below each color line correspond to $E_G-E'_G<0$ for a given $p$, i.e., the frozen electron state
represents the true ground state there. 
At $p=2$, the BCS state occupies the whole diagram,
whereas at $p\to \infty$, the border separating the two states coincides with the $\nu=\nu'$ line.

\section{Concluding remarks} 

Our results suggest that the key ingredient of our model is band flatness, which drastically amplifies interaction effects
and makes a transition into our special correlated state possible.  Moreover, we have shown that any negative effects of a small band gap or band asymmetry can always be eliminated through stronger band flatness. 

Recent studies of twisted graphene bilayers \cite{PRB-I,PRB-II,PRB-III,PRB-IV,PRB-V,PRB-VI} and multilayers \cite{TSTG-bernevig2021} reveal a versatile playground, which can potentially reproduce the flat bands we are after \cite{lau2022reproducibility}.
A possible experimental manifestation of our model would be an abrupt drop in resistivity upon decreasing electron temperature below a certain critical value, which resembles an enigmatic transition into a highly conductive (but not superconducting) state observed in twisted graphene multilayers at low temperatures \cite{he2021symmetry,xu2021tunable}. 
Hence, our model could be seen as a phenomenological one able to describe the phenomenon qualitatively.
The dimensionless pairing potential $\nu$ should be then considered a phenomenological parameter.
If the band flatness is substantial, then $\nu \sim \Delta/\epsilon_0$.
The band parameter $\epsilon_0$ can be seen as a bandwidth, which is up to several meV in magic-angle twisted multilayer graphene.
The critical temperature (the temperature at which resistivity drops down abruptly \cite{he2021symmetry})
is $\sim 10$ K which is equivalent to $\sim 1$ meV. The order parameter $\Delta$  at $T=0$ should be of the same order of magnitude
as the critical temperature. Hence, the phenomenological parameter $\nu$ is slightly below 1.

A somewhat closer realization of our model would be chirally stacked (rhombohedral) multilayer graphene with a number of layers $>2$ \cite{tarucha2009,macdonald2010,sopik2014charge,katsnelson2016}.
However, recent measurements performed in intrinsic rhombohedral-stacked pentalayer graphene indicate a correlated insulator
rather than a correlated conductor \cite{han2024correlated}.

We conclude with two caveats about our model.  First, the concept of frozen electrons is inspired by a mean-field theory.  As with any other mean-field theory, these results indicate just the possibility of the existence of frozen electrons but not a necessity.  Second, to transition into the frozen state, the flat-band electrons must be subject to mutual attraction, at least a weak one in the short range. The origin of attraction is not important; it could rely on one of the pairing mechanisms introduced in various superconductivity theories \cite{cea2021coulomb}. However, the intraband pairing must be weaker than the cross-band one;  otherwise, the electrons would just transition into the usual BCS state,
see Fig. \ref{fig5}.  It remains an open question for future work whether the special constraints required for our model can be realized by employing a specific pairing mechanism.

\acknowledgments
This research is supported by the Singapore Ministry of Education Research Centre of Excellence award to the Institute for Functional Intelligent Materials
(I-FIM, Project No. EDUNC-33-18-279-V12), and Singapore National Science Foundation Investigator Award (Grant No.~NRF-NRFI06-2020-0003).
M.T. acknowledges support from the Centre for Advanced 2D Materials funded within the Singapore National Science Foundation Medium Sized Centre Programme and
thanks Maksim Ulybyshev and Fakher Assaad for discussions and hospitality at the University of W\"urzburg. G.S. acknowledges support from SERB Grant No. SRG/2020/000134.

\appendix

\section{Quasiparticle scattering}
\label{appA}

 The quasiparticle excitations of $\gamma_{0,1\mathbf{p}}$ type at $|\mathbf{p}|\leq k_\Delta$ are given by
\begin{eqnarray}
 \gamma_{0\mathbf{p}}^\dagger\vert G \rangle & = & 0,\\
\nonumber \gamma_{0\mathbf{p}}\vert G \rangle & = & e^{i\delta} c_{-\mathbf{\bar p}\downarrow}^\dagger  
\prod\limits_{k\leq k_\Delta, \mathbf{k}\neq \mathbf{p}} \frac{1}{\sqrt{2}}\left(1 + e^{i\delta}c_{+\mathbf{k}\uparrow}^\dagger c_{-\mathbf{\bar k}\downarrow}^\dagger\right) \\
&&\times  \prod\limits_{k > k_\Delta} c_{-\mathbf{\bar k}\downarrow}^\dagger  \vert 0 \rangle \otimes |G\rangle_{\downarrow\uparrow},\\
\gamma_{1\mathbf{p}}^\dagger\vert G \rangle  & = & 0,\\
\nonumber 
\gamma_{1\mathbf{p}}\vert G \rangle & = &c_{+\mathbf{p}\uparrow}^\dagger \prod\limits_{k\leq k_\Delta, \mathbf{k}\neq \mathbf{p}} 
\frac{1}{\sqrt{2}}\left(1 +  e^{i\delta} c_{+\mathbf{k}\uparrow}^\dagger c_{-\mathbf{\bar k}\downarrow}^\dagger\right)\\
&& \times \prod\limits_{k > k_\Delta} c_{-\mathbf{\bar k}\downarrow}^\dagger \vert 0 \rangle \otimes |G\rangle_{\downarrow\uparrow}.
\end{eqnarray}
Here, like the conventional BCS model, each quasiparticle excitation
breaks a pair apart creating a free electron.
In contrast, the quasiparticle excitations outside of the frozen layer ($|\mathbf{p}|> k_\Delta$) read
\begin{eqnarray}
\nonumber  \gamma_{0\mathbf{p}}^\dagger\vert G \rangle & = & \frac{e^{-i\delta}}{\sqrt{2}}
\left(1 + e^{i\delta}c_{+\mathbf{p}\uparrow}^\dagger c_{-\mathbf{\bar p}\downarrow}^\dagger\right)
\prod\limits_{k > k_\Delta,\mathbf{k}\neq \mathbf{p}} c_{-\mathbf{\bar k}\downarrow}^\dagger \\
\nonumber  &&\times \prod\limits_{k\leq k_\Delta}  \frac{1}{\sqrt{2}}\left(1 + e^{i\delta}c_{+\mathbf{k}\uparrow}^\dagger c_{-\mathbf{\bar k}\downarrow}^\dagger\right)   \vert 0 \rangle \otimes |G\rangle_{\downarrow\uparrow},\\
\\
\gamma_{0\mathbf{p}}\vert G \rangle & = & 0,\\
\gamma_{1\mathbf{p}}^\dagger\vert G \rangle  & = & 0,\\
\nonumber 
\gamma_{1\mathbf{p}}\vert G \rangle & = &
-\frac{e^{-i\delta}}{\sqrt{2}}
\left(1 - e^{i\delta}c_{+\mathbf{p}\uparrow}^\dagger c_{-\mathbf{\bar p}\downarrow}^\dagger\right)
\prod\limits_{k > k_\Delta,\mathbf{k}\neq \mathbf{p}} c_{-\mathbf{\bar k}\downarrow}^\dagger \\
\nonumber  &&\times \prod\limits_{k\leq k_\Delta}  \frac{1}{\sqrt{2}}\left(1 + e^{i\delta}c_{+\mathbf{k}\uparrow}^\dagger c_{-\mathbf{\bar k}\downarrow}^\dagger\right)   \vert 0 \rangle \otimes |G\rangle_{\downarrow\uparrow}.\\
\end{eqnarray}
Here, each excitation creates an electron pair outside of the frozen layer.
Figure \ref{fig3}(c) illustrates the quasiparticle excitation processes.
The quasiparticle excitations of $\beta_{0,1\mathbf{p}}$ type can be obtained from the above relations by swapping the spin and momentum indices.

Each term in the disorder Hamiltonian given by Eq. (\ref{Hdis}) can be explicitly written
in terms of the quasiparticle operators as
\begin{eqnarray}
\label{line1}
  && \sum\limits_{\mathbf{kk}'} U_{\mathbf{kk'}} c_{+\mathbf{k}'\uparrow}^\dagger c_{+\mathbf{k}\uparrow} = \\
\nonumber   && \frac{1}{2}\sum\limits_{\mathbf{kk}'} U_{\mathbf{kk'}}\left(
 \gamma_{0\mathbf{k}'}^\dagger \gamma_{0\mathbf{k}}  + \gamma_{0\mathbf{k}'}^\dagger \gamma_{1\mathbf{k}}^\dagger
 + \gamma_{1\mathbf{k}'} \gamma_{0\mathbf{k}} + \gamma_{1\mathbf{k}'} \gamma_{1\mathbf{k}}^\dagger \right),\\
 && \sum\limits_{\mathbf{kk}'} U_{\mathbf{kk'}} c_{-\mathbf{\bar k}'\downarrow}^\dagger c_{-\mathbf{\bar k}\downarrow}  =\\
  \nonumber && \frac{1}{2}\sum\limits_{\mathbf{kk}'} U_{\mathbf{kk'}}\left(
 -\gamma_{0\mathbf{k}'}^\dagger \gamma_{0\mathbf{k}}  +  \gamma_{1\mathbf{k}'} \gamma_{0\mathbf{k}} 
 +  \gamma_{0\mathbf{k}'}^\dagger \gamma_{1\mathbf{k}}^\dagger -  \gamma_{1\mathbf{k}'} \gamma_{1\mathbf{k}}^\dagger \right),\\
 && \sum\limits_{\mathbf{kk}'} U_{\mathbf{kk'}} c_{+\mathbf{\bar k}'\downarrow}^\dagger c_{+\mathbf{\bar k}\downarrow}   = \\
 \nonumber   && \frac{1}{2}\sum\limits_{\mathbf{kk}'} U_{\mathbf{kk'}}\left(
 \beta_{0\mathbf{k}'}^\dagger \beta_{0\mathbf{k}}  + \beta_{0\mathbf{k}'}^\dagger \beta_{1\mathbf{k}}^\dagger
 + \beta_{1\mathbf{k}'} \beta_{0\mathbf{k}} + \beta_{1\mathbf{k}'} \beta_{1\mathbf{k}}^\dagger \right),\\
   &&  \sum\limits_{\mathbf{kk}'} U_{\mathbf{kk}'} c_{-\mathbf{k}'\uparrow}^\dagger c_{-\mathbf{k}\uparrow} = \\
 \nonumber   && \frac{1}{2}\sum\limits_{\mathbf{kk}'} U_{\mathbf{kk'}} \left(
 - \beta_{0\mathbf{k}'}^\dagger\beta_{0\mathbf{k}}  + \beta_{1\mathbf{k}'}\beta_{0\mathbf{k}} 
 +  \beta_{0\mathbf{k}'}^\dagger\beta_{1\mathbf{k}}^\dagger -  \beta_{1\mathbf{k}'} \beta_{1\mathbf{k}}^\dagger \right),\\
 && \sum\limits_{\mathbf{kk}'} U_{\mathbf{kk}'} c_{+\mathbf{k}'\uparrow}^\dagger c_{-\mathbf{k}\uparrow}  = \\
 \nonumber   && \frac{e^{i\delta}}{2}\sum\limits_{\mathbf{kk}'} U_{\mathbf{kk}'}\left(
 \gamma_{0\mathbf{k}'}^\dagger \beta_{0\mathbf{k}}^\dagger  - \gamma_{0\mathbf{k}'}^\dagger \beta_{1\mathbf{k}}
 + \gamma_{1\mathbf{k}'} \beta_{0\mathbf{k}}^\dagger - \gamma_{1\mathbf{k}'} \beta_{1\mathbf{k}}\right),\\
 &&  \sum\limits_{\mathbf{kk}'} U_{\mathbf{kk}'} c_{+\mathbf{\bar k}'\downarrow}^\dagger c_{-\mathbf{\bar k}\downarrow}  = \\
 \nonumber   &&  \frac{e^{i\delta}}{2}\sum\limits_{\mathbf{kk}'} U_{\mathbf{kk}'}\left(
 -\gamma_{0\mathbf{k}'}^\dagger \beta_{0\mathbf{k}}^\dagger  +  \gamma_{1\mathbf{k}'}\beta_{0\mathbf{k}}^\dagger 
 - \gamma_{0\mathbf{k}'}^\dagger\beta_{1\mathbf{k}} +  \gamma_{1\mathbf{k}'} \beta_{1\mathbf{k}}\right),\\
  && \sum\limits_{\mathbf{kk}'} U_{\mathbf{kk}'} c_{-\mathbf{k}'\uparrow}^\dagger c_{+\mathbf{k}\uparrow} = \\
  \nonumber   && \frac{e^{-i\delta}}{2}\sum\limits_{\mathbf{kk}'} U_{\mathbf{kk}'}\left(
 \beta_{0\mathbf{k}'} \gamma_{0\mathbf{k}}  + \beta_{0\mathbf{k}'} \gamma_{1\mathbf{k}}^\dagger
 - \beta_{1\mathbf{k}'}^\dagger \gamma_{0\mathbf{k}} - \beta_{1\mathbf{k}'}^\dagger \gamma_{1\mathbf{k}}^\dagger \right),\\
  \label{line8} 
  && \sum\limits_{\mathbf{kk}'} U_{\mathbf{kk}'} c_{-\mathbf{\bar k}'\downarrow}^\dagger c_{+\mathbf{\bar k}\downarrow}  = \\
 \nonumber   && \frac{e^{-i\delta}}{2}\sum\limits_{\mathbf{kk}'} U_{\mathbf{kk}'}\left(
 - \beta_{0\mathbf{k}'}\gamma_{0\mathbf{k}}  -  \beta_{1\mathbf{k}'}^\dagger \gamma_{0\mathbf{k}}
 +  \beta_{0\mathbf{k}'}\gamma_{1\mathbf{k}}^\dagger +  \beta_{1\mathbf{k}'}^\dagger \gamma_{1\mathbf{k}}^\dagger\right).
\end{eqnarray}
Here, we have assumed $U_{\mathbf{k}'\mathbf{k}}=U_{\mathbf{kk}'}$.
Each term in Eqs. (\ref{line1})--(\ref{line8}) describes a certain scattering channel
contributing to the total momentum relaxation.
Summing up Eqs. (\ref{line1})--(\ref{line8}), we obtain Eq. (\ref{Hdis2}).

The only nonzero scattering matrix elements are 
\begin{eqnarray}
\nonumber\langle G \vert  \gamma_{1\mathbf{p}'}^\dagger H_\mathrm{dis} \gamma_{0\mathbf{p}}^\dagger \vert G \rangle 
\nonumber& = & \sum\limits_{\mathbf{kk}'} U_{\mathbf{kk}'}\langle G \mid  \gamma_{1\mathbf{p}'}^\dagger \gamma_{1\mathbf{k}'} \gamma_{0\mathbf{k}} \gamma_{0\mathbf{p}}^\dagger \mid G \rangle \\
\nonumber &= &  \sum\limits_{\mathbf{kk}'}  \left\{ \begin{array}{ll}
   U_{\mathbf{kk}'}\delta_{\mathbf{p}'\mathbf{k}'}\delta_{\mathbf{p}\mathbf{k}}, & |\mathbf{p}|>k_\Delta, \\  0, & \text{otherwise} .   \end{array} \right. \\
& = & \left\{ \begin{array}{ll}
   U_{\mathbf{pp}'}, & |\mathbf{p}|>k_\Delta, \\  0, & \text{otherwise}.   \end{array} \right.
   \label{gamma1}
\end{eqnarray}
\begin{eqnarray}
\nonumber\langle G \vert  \gamma_{0\mathbf{p}'} H_\mathrm{dis} \gamma_{1\mathbf{p}} \vert G \rangle 
\nonumber& = & \sum\limits_{\mathbf{kk}'} U_{\mathbf{kk}'}\langle G \mid  \gamma_{0\mathbf{p}'} \gamma_{0\mathbf{k}'}^\dagger \gamma_{1\mathbf{k}}^\dagger \gamma_{1\mathbf{p}} \mid G \rangle \\
\nonumber &= &  \sum\limits_{\mathbf{kk}'}  \left\{ \begin{array}{ll}
   U_{\mathbf{kk}'}\delta_{\mathbf{p}'\mathbf{k}'}\delta_{\mathbf{p}\mathbf{k}}, & |\mathbf{k}'|>k_\Delta, \\  0, & \text{otherwise} .   \end{array} \right. \\
& = & \left\{ \begin{array}{ll}
   U_{\mathbf{pp}'}, & |\mathbf{p}'|>k_\Delta, \\  0, & \text{otherwise}.   \end{array} \right.
   \label{gamma0}
\end{eqnarray}
\begin{eqnarray}
\nonumber \langle G \vert  \beta_{1\mathbf{p}'}^\dagger H_\mathrm{dis} \beta_{0\mathbf{p}}^\dagger \vert G \rangle 
\nonumber & = & \sum\limits_{\mathbf{kk}'} U_{\mathbf{kk}'}\langle G \mid  \beta_{1\mathbf{p}'}^\dagger \beta_{1\mathbf{k}'} \beta_{0\mathbf{k}} \beta_{0\mathbf{p}}^\dagger \mid G \rangle \\
\nonumber &= &  \sum\limits_{\mathbf{kk}'}  \left\{ \begin{array}{ll}
   U_{\mathbf{kk}'}\delta_{\mathbf{p}'\mathbf{k}'}\delta_{\mathbf{p}\mathbf{k}}, & |\mathbf{p}|>k_\Delta, \\  0, & \text{otherwise} .   \end{array} \right. \\
& = & \left\{ \begin{array}{ll}
   U_{\mathbf{pp}'}, & |\mathbf{p}|>k_\Delta, \\  0, & \text{otherwise}.   \end{array} \right.
\label{beta1}
\end{eqnarray}
\begin{eqnarray}
\nonumber  \langle G \vert  \beta_{0\mathbf{p}'} H_\mathrm{dis} \beta_{1\mathbf{p}} \vert G \rangle 
\nonumber  & = & \sum\limits_{\mathbf{kk}'} U_{\mathbf{kk}'}\langle G \mid  \beta_{0\mathbf{p}'} \beta_{0\mathbf{k}'}^\dagger \beta_{1\mathbf{k}}^\dagger \beta_{1\mathbf{p}} \mid G \rangle \\
\nonumber  &= &  \sum\limits_{\mathbf{kk}'}  \left\{ \begin{array}{ll}
   U_{\mathbf{kk}'}\delta_{\mathbf{p}'\mathbf{k}'}\delta_{\mathbf{p}\mathbf{k}}, & |\mathbf{k}'|>k_\Delta, \\  0, & \text{otherwise} .   \end{array} \right. \\
& = & \left\{ \begin{array}{ll}
   U_{\mathbf{pp}'}, & |\mathbf{p}'|>k_\Delta, \\  0, & \text{otherwise}.   \end{array} \right.
   \label{beta0}
\end{eqnarray}

\begin{eqnarray}
\nonumber  \langle G \vert  \gamma_{1\mathbf{p}'}^\dagger H_\mathrm{dis} \beta_{0\mathbf{p}} \vert G \rangle 
\nonumber  & = & \sum\limits_{\mathbf{kk}'} U_{\mathbf{kk}'}e^{i\delta}\langle G \mid  \gamma_{1\mathbf{p}'}^\dagger \gamma_{1\mathbf{k}'} \beta_{0\mathbf{k}}^\dagger \beta_{0\mathbf{p}} \mid G \rangle \\
\nonumber  &= &  \sum\limits_{\mathbf{kk}'}  \left\{ \begin{array}{ll}
   U_{\mathbf{kk}'}e^{i\delta}\delta_{\mathbf{p}'\mathbf{k}'}\delta_{\mathbf{p}\mathbf{k}}, & |\mathbf{p}|<k_\Delta, \\  0, & \text{otherwise} .   \end{array} \right. \\
& = & \left\{ \begin{array}{ll}
   U_{\mathbf{pp}'}e^{i\delta}, & |\mathbf{p}|<k_\Delta, \\  0, & \text{otherwise}.   \end{array} \right.
   \label{01}
\end{eqnarray}
\begin{eqnarray}
\nonumber \langle G \vert  \beta_{0\mathbf{p}'}^\dagger H_\mathrm{dis} \gamma_{1\mathbf{p}} \vert G \rangle 
\nonumber & = & \sum\limits_{\mathbf{kk}'} U_{\mathbf{kk}'}e^{-i\delta}\langle G \mid  \beta_{0\mathbf{p}'}^\dagger \beta_{0\mathbf{k}'}\gamma_{1\mathbf{k}}^\dagger \gamma_{1\mathbf{p}} \mid G \rangle \\
\nonumber &= &  \sum\limits_{\mathbf{kk}'}  \left\{ \begin{array}{ll}
   U_{\mathbf{kk}'} e^{-i\delta}\delta_{\mathbf{p}'\mathbf{k}'}\delta_{\mathbf{p}\mathbf{k}}, & |\mathbf{k}'|<k_\Delta, \\  0, & \text{otherwise} .   \end{array} \right. \\
& = & \left\{ \begin{array}{ll}
   U_{\mathbf{pp}'}e^{-i\delta}, & |\mathbf{p}'|<k_\Delta, \\  0, & \text{otherwise}.   \end{array} \right.
   \label{10}
\end{eqnarray}

\begin{eqnarray}
\nonumber \langle G \vert  \gamma_{0\mathbf{p}'}^\dagger H_\mathrm{dis} \beta_{1\mathbf{p}} \vert G \rangle 
\nonumber & = & \sum\limits_{\mathbf{kk}'} U_{\mathbf{kk}'}e^{-i\delta}\langle G \mid  \gamma_{0\mathbf{p}'}^\dagger  \gamma_{0\mathbf{k}'} \beta_{1\mathbf{k}}^\dagger \beta_{1\mathbf{p}} \mid G \rangle \\
\nonumber &= &  \sum\limits_{\mathbf{kk}'}  \left\{ \begin{array}{ll}
   U_{\mathbf{kk}'}e^{-i\delta}\delta_{\mathbf{p}'\mathbf{k}'}\delta_{\mathbf{k}\mathbf{p}}, & |\mathbf{k}'|<k_\Delta, \\  0, & \text{otherwise} .   \end{array} \right. \\
& = & \left\{ \begin{array}{ll}
   U_{\mathbf{pp}'}e^{-i\delta}, & |\mathbf{p}'|<k_\Delta, \\  0, & \text{otherwise}.   \end{array} \right.
   \label{01spin}
\end{eqnarray}
\begin{eqnarray}
\nonumber \langle G \vert  \beta_{1\mathbf{p}'}^\dagger H_\mathrm{dis} \gamma_{0\mathbf{p}} \vert G \rangle 
\nonumber & = & \sum\limits_{\mathbf{kk}'} U_{\mathbf{kk}'}e^{i\delta}\langle G \mid  \beta_{1\mathbf{p}'}^\dagger  \beta_{1\mathbf{k}'} \gamma_{0\mathbf{k}}^\dagger  \gamma_{0\mathbf{p}} \mid G \rangle \\
\nonumber &= &  \sum\limits_{\mathbf{kk}'}  \left\{ \begin{array}{ll}
   U_{\mathbf{kk}'}e^{i\delta}\delta_{\mathbf{p}'\mathbf{k}'}\delta_{\mathbf{p}\mathbf{k}}, & |\mathbf{p}|<k_\Delta, \\  0, & \text{otherwise} .   \end{array} \right. \\
& = & \left\{ \begin{array}{ll}
   U_{\mathbf{pp}'}e^{i\delta}, & |\mathbf{p}|<k_\Delta, \\  0, & \text{otherwise}.   \end{array} \right.
   \label{10spin}
\end{eqnarray} 
In the case of short-range ($\delta$-shaped) scatterers, the  scattering matrix elements are constants ($U_{\mathbf{pp}'}=U_0$),
and the momentum relaxation rate can be written as
\begin{eqnarray}
 \nonumber \frac{1}{\tau_{\kappa k}}=&&\frac{2\pi}{\hbar}\sum\limits_{\kappa'}\int\frac{d^2 k'}{(2\pi)^2} n_\mathrm{dis}
 \left\vert\langle \kappa \mathbf{k} \vert H_\mathrm{dis} \vert \kappa' \mathbf{k}' \rangle\right\vert|^2\\
 && \times \delta\left(E_{\kappa k}-E_{\kappa'k'}\right),
 \label{tau}
\end{eqnarray}
where $n_\mathrm{dis}$ is the disorder concentration, $\kappa$ denotes the quasiparticle type $\{\gamma_{0,1}, \beta_{0,1}\}$,
and $E_{\kappa k}$ is the quasiparticle excitation energy shown in Fig. \ref{fig3}(c).
Equation (\ref{tau}) involves summation over the quasiparticle scattering channels specified
by Eqs. (\ref{gamma1})--(\ref{10spin}). The equations demonstrate that scattering is only possible
between the quasiparticles of opposite type, e.g., between $\gamma_{0\mathbf{k}}$ 
and $\gamma_{1\mathbf{k}}$ (or $\gamma_{0\mathbf{k}}$ and  $\beta_{1\mathbf{k}}$) but
never between $\gamma_{0\mathbf{k}}$  and $\beta_{0\mathbf{k}}$ (or $\gamma_{1\mathbf{k}}$ and  $\beta_{1\mathbf{k}}$).
Since $\gamma$ and $\beta$ quasiparticles have the same energy spectrum,
the delta function takes the same form for each scattering channel given by
\begin{eqnarray}
\nonumber \delta\left(E_{\kappa k}-E_{\kappa'k'}\right) & = &\delta(\epsilon_{k}+\epsilon_{k'})\\
& = & 0, \quad \epsilon_{k,k'}>0.
\end{eqnarray}
The later equality results in $\tau_{\kappa k} = \infty$.

It is instructive to calculate the momentum relaxation rate in the normal state.
One can make use of Eq. (\ref{tau}) assuming that $E_{\kappa k}=\kappa \epsilon_k$,
where $\kappa=\pm$ is the band index.
If electrons are in the normal state, then the intraband scattering channels are allowed; hence, the integral does not vanish, and the momentum relaxation rate reads
\begin{equation}
 \frac{1}{\tau_k} = \frac{n_\mathrm{dis} U_0^2 k_0^2}{p\hbar\epsilon_0}\left(\frac{k_0}{k}\right)^{p-2}.
\end{equation}
One can now see the difference between electron scattering in the normal and correlated states:
Intraband elastic scattering is allowed for normal electrons but can be suppressed by electron pairing,
whereas interband elastic scattering is forbidden regardless.

\bibliography{collinear.bib,myBCS.bib}

\end{document}